# CO Oxidation Catalysed by Pd-based Bimetallic Nanoalloys

**Dennis Palagin,**[*a] **and Jonathan P. K. Doye**[a]

Density functional theory based global geometry optimization has been used to demonstrate the crucial influence of the geometry of the catalytic cluster on the energy barriers for the CO oxidation reaction over Pd-based bimetallic nanoalloys. We show that dramatic geometry change between the reaction intermediates can lead to very high energy barriers and thus be prohibitive for the whole process. This introduces challenges for both the design of new catalysts, and theoretical methods employed. On the theory side, a careful choice of geometric configurations of all reaction intermediates is crucial for an adequate description of a possible reaction path. From the point of view of the catalyst design, the cluster geometry can be controlled by adjusting the level of interaction between the cluster and the dopant metal, as well as between the adsorbate molecules and the catalyst cluster by mixing different metals in a single nanoalloy particle. We show that substitution of a Pd atom in the $Pd_5$ cluster with a single Ag atom to form $Pd_4Ag_1$ leads to a potential improvement of the catalytic properties of the cluster for the CO oxidation reaction. On the other hand, a single Au atom does not enhance the properties of the catalyst, which is attributed to a weaker hybridization between the cluster's constituent metals and the adsorbate molecules. Such flexibility of properties of bimetallic nanoalloy clusters illustrates the possibility of fine-tuning, which might be used for design of novel efficient catalytic materials.

## 1 Introduction

Understanding catalysis at the atomic level is one of the fundamental goals of chemistry. Supported metal cluster catalysis in particular plays an enormous role in the modern chemical industry, and much effort is devoted to understanding this phenomenon and to the design and development of metal-cluster-based heterogeneous catalysts with high activity, selectivity, and stability.[1] Bimetallic catalysts often exhibit advantageous properties compared to those of their pure constituent metals, as this hybrid material may offer opportunities for the synergistic inter-metallic interactions that lead to improvement in catalytic performance.[2] Moreover, the catalytic activity of bimetallic alloys or nanoparticles can be optimized by controlling structural factors, such as the alloying element and concentration.[3] The geometric and electronic effects have been shown to systematically alter the catalytic activity of bimetallic catalysts.[4]

For example, it has been found that the addition of metals that bind oxygen strongly (e.g. Co, Ni, and Cu) to more noble metals (Pt or Pd) can improve their oxygen reduction catalytic activity.[5–8] Bimetallic clusters are also known to be effective catalysts for methane dehydrogenation,[9] and a rich variety of organic reactions, such as the Suzuki reaction,[10] the dehalogenation of aromatic cycles,[11,12] and the oxidation of benzyl alcohol,[13] among others.

[a] *Physical and Theoretical Chemistry Laboratory, Department of Chemistry, University of Oxford, South Parks Road, Oxford, OX1 3QZ, United Kingdom. E-mail: dennis.palagin@chem.ox.ac.uk*

A very important process catalysed by bimetallic clusters is the CO oxidation reaction, which has a large impact in terms of environmental protection. Great efforts have been devoted to developing highly efficient catalysts to remove the poisonous CO gas from car exhaust and from hydrogen gases used for fuel cells.[14] Theoretical studies of the CO adsorption and oxidation have been carried out for Au–Pd,[15] Pd–Pt, Cu–Pt, Pd–Rh,[16] Zr–Sc,[17] and Au–Cu[18] clusters. Experimentally, catalytic CO oxidation over Pd–Ni has been reported recently.[19] Theoretical works propose both Langmuir-Hinshelwood (two molecules adsorb and the adsorbed molecules undergo a bimolecular reaction) and Eley-Rideal (one of the molecules adsorbs and the other one reacts with it directly from the gas phase) mechanisms. For example, in the case of small Au–Pd clusters, both reaction mechanisms were investigated, with the Langmuir-Hinshelwood found to be more energetically favourable.[15]

If there is oxygen present in the system, such as in the case of Zr–Sc oxide clusters, oxygen atoms for CO oxidation are provided by the cluster.[17] In the case of Au–Cu clusters adsorbed on the oxide surface, the co-adsorption of CO and $O_2$ (Langmuir-Hinshelwood) takes place.[18] Thus, the Langmuir-Hinshelwood mechanism seems to be preferable for CO oxidation on bimetallic clusters. Another concern is the stability of the catalyst upon adsorption. The data in Ref. 16 indicates the possibility of structural rearrangements in the Au–Pd clusters upon adsorption of the CO.

Theoretical methods are therefore especially suited for studying the catalytic properties of such complex materials,

as they allow one to obtain detailed information on the mechanism of the reaction and the structural rearrangements involved, which is not directly accessible to experiment. Moreover, theory can aid the design of a novel catalyst by testing the catalytic activity of the systems of variable size and composition against a suitable performance descriptor.[20]

A number of theoretical studies have been conducted to elucidate the influence of the dopant metals on the catalytic activity of clusters.[21,22] It has been suggested that substitution of a single Pd atom in a $Pd_5$ cluster with a Au atom might improve the catalytic activity.[21] Moreover, it has been suggested that adsorption of such a $Pd_4Au$ cluster on a $TiO_2$ (110) surface leads to a more efficient CO to $CO_2$ conversion, compared to a pure $Pd_5$ cluster. At the same time, pure gold clusters have also been recently proposed to potentially exhibit high catalytic activity for the same process, e.g. $Au_8$ adsorbed on ZnO.[22] Thus the question of the optimal ratio of Pd and Au atoms in the catalytic Pd–Au clusters remains unsolved.

A very important open problem is finding the most relevant configurations of the intermediates at every stage of the reaction. The reaction path obviously depends on the cluster configuration, as concominantly do the heights of the energy barriers. Ideally one would like to characterize the complete potential energy surfaces for clusters and adsorbates to identify the most likely pathway for the catalytic reaction. However, due to the expense of the level of theory needed to obtain reliable results and the large number of different structures one has to consider for alloy clusters, this is potentially not feasible. This is the main bottleneck of a theoretical investigation of catalysis on nanoalloys. What is the best approach in the absence of a complete characterization of a potential energy surface is an open question.

Often the catalyst is assumed stable during the reaction, and only small adsorbate molecules (such as $O_2$ and CO in our case) are allowed to undergo major configurational changes, while the geometry of the catalytic metal cluster is only relaxed locally. This assumes that the rearrangement of the cluster geometry is slow compared to the overall rate of catalytic reaction. Assuming such stability of the cluster is likely to lead to relatively simple reaction paths and perhaps relatively low transition state energy barriers. However, if there is sufficient energy and time to allow significant reconfigurations in the adsorbate geometry, why should the cluster geometry not relax to more stable geometries if available?

An alternative approach is to assume that the cluster adopts the ground-state configuration at every step in the reaction. This approach assumes that the rearrangement of the cluster is relatively fast compared to the overall rate of the catalytic reaction, and requires a global optimization of the geometry at every step. To validate the applicability of this approach, such a "ground-state-based" reaction path should, of course, be compared to other possible paths involving low-lying isomers of every intermediate.

These open questions motivate the present theoretical study of the CO oxidation catalysed by Pd-based bimetallic nanoalloys. We first find a reasonable Pd/dopant ratio based on the adsorption energies of CO and $O_2$ molecules on the corresponding clusters. After that we study the complete reaction path of the CO oxidation for the Pd clusters alloyed with all possible dopants from the typically catalytically active groups VIIIB and IB: Au, Ag, Cu, Ni, Pt. Here, we rely on the thermodynamic feasibility of the catalytic process, thus effectively using the heights of the energy barriers along the reaction path as a descriptor of whether the catalyst is potentially active. Finally, we discuss the observed differences in the behaviour of the clusters of different composition based on an analysis of the electronic structure.

## 2 Computational details

All local geometry optimizations of the discussed structures, and subsequent electronic structure analysis were carried out with the plane-wave DFT package `CASTEP`.[23] Electronic exchange and correlation was treated within the generalized-gradient approximation functional due to Perdew, Burke and Ernzerhof (PBE).[24] The core electrons were described using ultrasoft pseudopotentials, whereas the valence electrons were treated with a plane-wave basis set with a cut-off energy of 300 eV. Local structure optimization is done using the Broyden-Fletcher-Goldfarb-Shanno method,[25] relaxing all force components to smaller than $10^{-2}$ eV/Å. The stability of the identified minima has been confirmed by vibrational frequency analysis.

To obtain the ground-state structures for all considered systems, we relied on basin-hopping (BH) based global geometry optimization,[26,27] using the DFT total energies and atomic forces calculated by `CASTEP`[23] as implemented in the `Atomic Simulation Environment (ASE)` suite.[28]

The Nudged Elastic Band (NEB)[29] method as implemented in the `ASE`[28] package was used to find the transition paths and corresponding energy barriers between given initial and final states, which were defined as the ground-state structures of adjacent reaction intermediates. Permutational isomers of the global minima were chosen in such a way as to minimize the geometry change involved in the transition. A chain of twelve replicas of the system has been constructed for every transition state calculation, and the forces were allowed to relax with a threshold of 0.05 eV. The two states are assumed to be connected by a single barrier path. For the transitions involving the loss of $CO_2$ the corresponding final minimum is reoptimized with a desorbed $CO_2$ molecule explicitly present in the system.

The reaction path of CO oxidation over a bimetallic catalyst is defined as

$$2CO + O_2 + Pd_xM_y \rightarrow 2CO_2 + Pd_xM_y,$$

where $M$ is a doping metal atom. The possible intermediate products are assumed as CO–$Pd_xM_y$, $O_2$–$Pd_xM_y$, $O_2$CO–$Pd_xM_y$, O–$Pd_xM_y$, and OCO–$Pd_xM_y$. The total exothermic effect of the reaction is computed to be 6.41 eV (147.8 kcal/mol), with the experimental value reported as 135 kcal/mol.[30]

## 3 Selecting suitable cluster size and Pd/dopant ratio

In order to identify a system that is likely to have good catalytic properties, we suggest that a candidate catalyst should meet the following requirements:

- Adsorption of CO and $O_2$ is reasonably strong and of similar strength.
- Dissociative adsorption of $O_2$, or a substantial destabilization of the O$-$O bond is observed.
- Reactions of the adsorbed species on the cluster are characterized by low barriers.
- All the intermediates are less stable than the products of the reaction.
- The cluster shows little affinity for $CO_2$.

The first issue to address is to find a suitable size of the cluster, and a suitable ratio of atom types in the cluster. We use the first two requirements from the above list to assess different clusters. Ref. 21 suggested a catalytically active $Pd_5$ cluster as a reference point for Pd to Au substitution. Following this logic, we decided to either substitute atoms in such a cluster, thus ending up with a set of six $Pd_{5-x}Au_x$ clusters, or to cap it with an additional atom, thus having a set of seven $Pd_{6-x}Au_x$ clusters of various composition. We started with Pd–Au clusters as a model system, and later introduced other transition metal dopants for nanoalloys with the selected Pd/dopant ratio that was thought most likely to yield good catalytic properties.

As can be seen in Fig. 1, clusters having more than two gold atoms have a tendency to prefer planar geometries as their ground states. Thus, $Au_5$, $Pd_1Au_4$, $Au_6$, and $Pd_1Au_5$ are planar both as individual clusters, and with either CO or $O_2$ adsorbed. This is in line with the known tendency of small gold clusters to retain planar configurations up to the size of at least $Au_7$,[31] and at even larger sizes in the case of anionic clusters.[32] $Pd_2Au_3$ and $Pd_2Au_4$ clusters turn into planar structures upon adsorption of CO or $O_2$. This can be explained in terms of the reduced hybridization between the states of Pd and Au, when the Pd atoms predominantly interact with an adsorbate molecule rather than with the neighbouring Au atoms. Indeed, rather strong interactions between the adsorbates and the Pd atoms of the cluster (up to 2.49 eV binding energy in the case of the CO–$Pd_2Au_3$ aggregate) allow the relatively undisturbed gold atoms to retain the relativistic aspects[33] of their properties. This is also reflected in the tendency of gold atoms to cluster together in these cases. Another interesting feature worth mentioning is the fact that both CO and $O_2$ tend to directly interact with Pd atoms, and not Au atoms. The $Pd_3Au_3$ cluster can be qualified as an intermediate case, showing a planar configuration for the CO–$Pd_3Au_3$ aggregate, and a compact geometry for the $O_2$–$Pd_3Au_3$ aggregate.

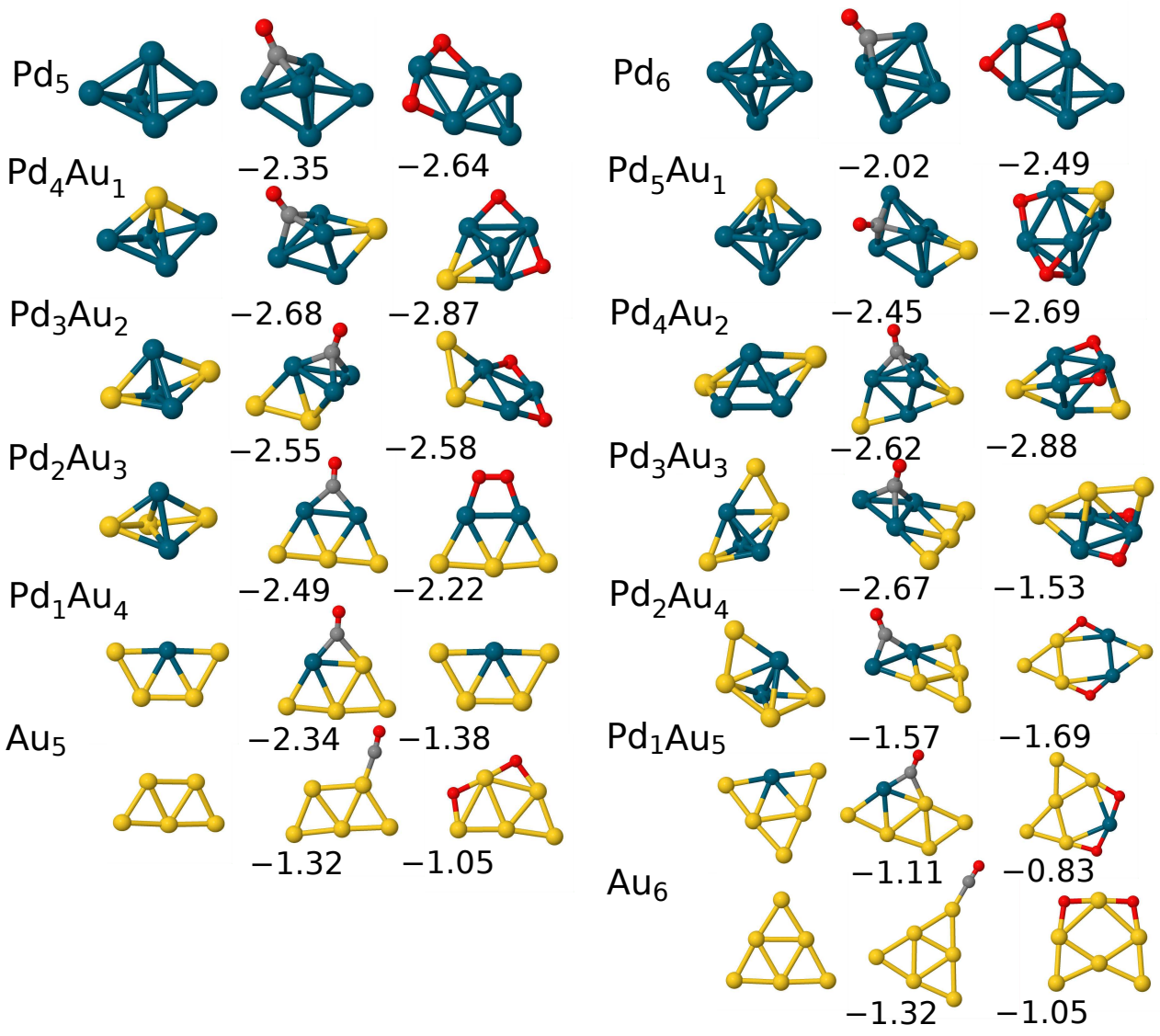

**Fig. 1** Lowest-energy $Pd_{5-x}Au_x$ and $Pd_{6-x}Au_x$ clusters, and their aggregates with CO and $O_2$. Numbers indicate adsorption energies in eV.

Clusters with a fraction of gold atoms that is less than half possess compact structures, and, more importantly, have higher binding energies towards both CO and $O_2$. In certain cases (such as $Pd_2Au_3$ and $Pd_1Au_4$) the $O_2$ molecule is not even dissociated upon adsorption. Thus a sufficient fraction of Pd atoms allows more efficient activation of the adsorbate molecules for the oxidation reaction, while compact structures ensure the stability of the cluster itself under realistic conditions, such as deposition on a support surface. From analysis of the CO and $O_2$ adsorption energies we conclude that $Pd_4Au_1$ and $Pd_4Au_2$ represent particularly good compositions in terms of the adsorption efficiency: these cluster sizes have compact geometries, ensure the dissociation of the $O_2$ molecule, and have the highest binding energies towards CO and $O_2$ among all considered clusters.

This conclusion is also backed up by similar calculations performed for the clusters deposited on the $TiO_2$ (110) surface, which is a widely used support for catalytic metal clusters in the reaction of CO oxidation.[34–36] $Pd_4Au_1$ and $Pd_4Au_2$ provide the highest adsorption energies for both CO and $O_2$ molecules compared to other Pd/Au ratios, except for pristine Pd clusters: deposited $Pd_5$ and $Pd_6$ yield the highest adsorption energies in all cases. Fig. 2 depicts the ground-state structures for the pure clusters and their aggregates with CO and $O_2$ in the cases of $Pd_4Au_1$ and $Pd_4Au_2$. For comparison, the corresponding structures of pristine $Pd_5$ and $Pd_6$ clusters are also provided.

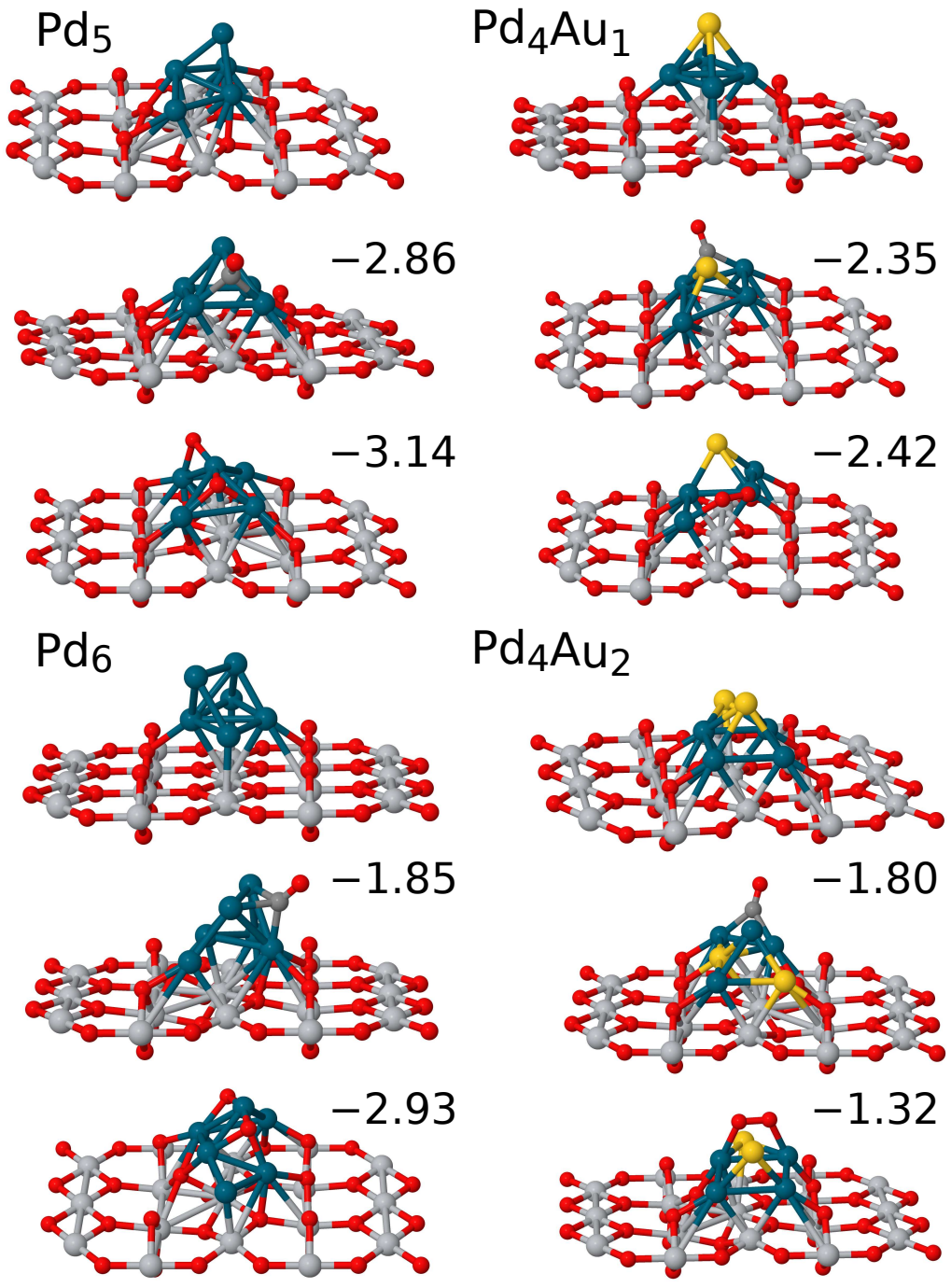

**Fig. 2** Lowest-energy $Pd_5$, $Pd_4Au_1$, $Pd_6$, and $Pd_5Au_1$ clusters adsorbed on the $TiO_2$ (110) surface, and their aggregates with CO and $O_2$. Only the top layer of the $TiO_2$ support is displayed for clarity. Numbers indicate adsorption energies in eV. See supplemental material[37] for the structures and energies of other Pd/Au compositions.

Interestingly, the oxygen atoms in the top layer of the $TiO_2$ (110) surface can participate in the stabilization of the clusters with adsorbed molecules, as can be seen by the change in their positions compared to the clean surface. By contrast, bare clusters retain their stability upon deposition, and do not disturb the surface much. However, in the presence of the adsorbate molecules the picture is quite different. While the ground-state geometries of the CO–$Pd_5$ and CO–$Pd_4Au_1$ aggregates are very similar to their pure cluster counterparts, the addition of the CO adsorbate molecule changes the global minima configurations of the deposited $Pd_6$ and $Pd_4Au_2$ clusters significantly. The adsorption of the $O_2$ molecule leads to rather distorted final structures in all cases. Even more striking is the fact that the addition of gold atoms prevents dissociation of the oxygen molecule. For both $Pd_4Au_1$ and $Pd_4Au_2$ clusters, the ground-state structures of the aggregates with oxygen correspond to activation of the $O_2$ molecule (O–O bond length of 1.32 Å and 1.35 Å in the case of $Pd_4Au_1$ and $Pd_4Au_2$, respectively, compared to 1.21 Å for a gas-phase $O_2$ molecule), but not dissociation. This behaviour can be explained by the observed partial charge transfer from the deposited cluster towards the surface. The resulting electron deficit prevents the cluster from becoming an active site for the dissociation of the adsorbed $O_2$ molecule which has a high electron affinity. However, previous theoretical studies[21,38–40] suggest that this fact does not necessarily prevent the $TiO_2$ (110)-deposited clusters from exhibiting efficient catalytic properties.

## 4 Selecting suitable dopants

Thus in our further investigations we focus on the cluster of $Pd_4M_1$ and $Pd_4M_2$. As dopant atoms we have chosen the metals from the typically catalytic groups VIIIB and IB, i.e. Au, Ag, Cu, Ni, and Pt. For comparison, pristine $Pd_5$ and $Pd_6$ clusters were also considered. In order to investigate the detailed mechanism of the reaction we initially take the "global minimum approach" where we assume all intermediates adopt their lowest-energy structure. Consequently, we run global geometry optimization for every intermediate of the assumed reaction path, i.e. $Pd_xM_y$, CO–$Pd_xM_y$, $O_2$–$Pd_xM_y$, $O_2$CO–$Pd_xM_y$, O–$Pd_xM_y$, and OCO–$Pd_xM_y$. Transition state structures were found by applying the NEB method using the ground-state structures of the adjacent intermediates as initial and final configurations, respectively. Thus the first transition state (TS1) corresponds to the reaction path between the $O_2$CO–$Pd_xM_y$ and O–$Pd_xM_y$ with a first $CO_2$ molecule released. The second transition state (TS2) corresponds to the reaction path between OCO–$Pd_xM_y$ and an individual cluster with a second $CO_2$ molecule released. Note that the final configuration of the cluster (after two $CO_2$ molecules have been produced) is assumed to match the initial ground-state structure of the cluster. In the following we first consider the reaction paths for $Pd_5$-based and $Pd_6$-based clusters separately, and then discuss the common features.

### 4.1 Reaction paths for $Pd_5$-based clusters

The identified reaction paths for the $Pd_5$-based clusters are presented in Fig. 3. The reaction path for the pristine $Pd_5$

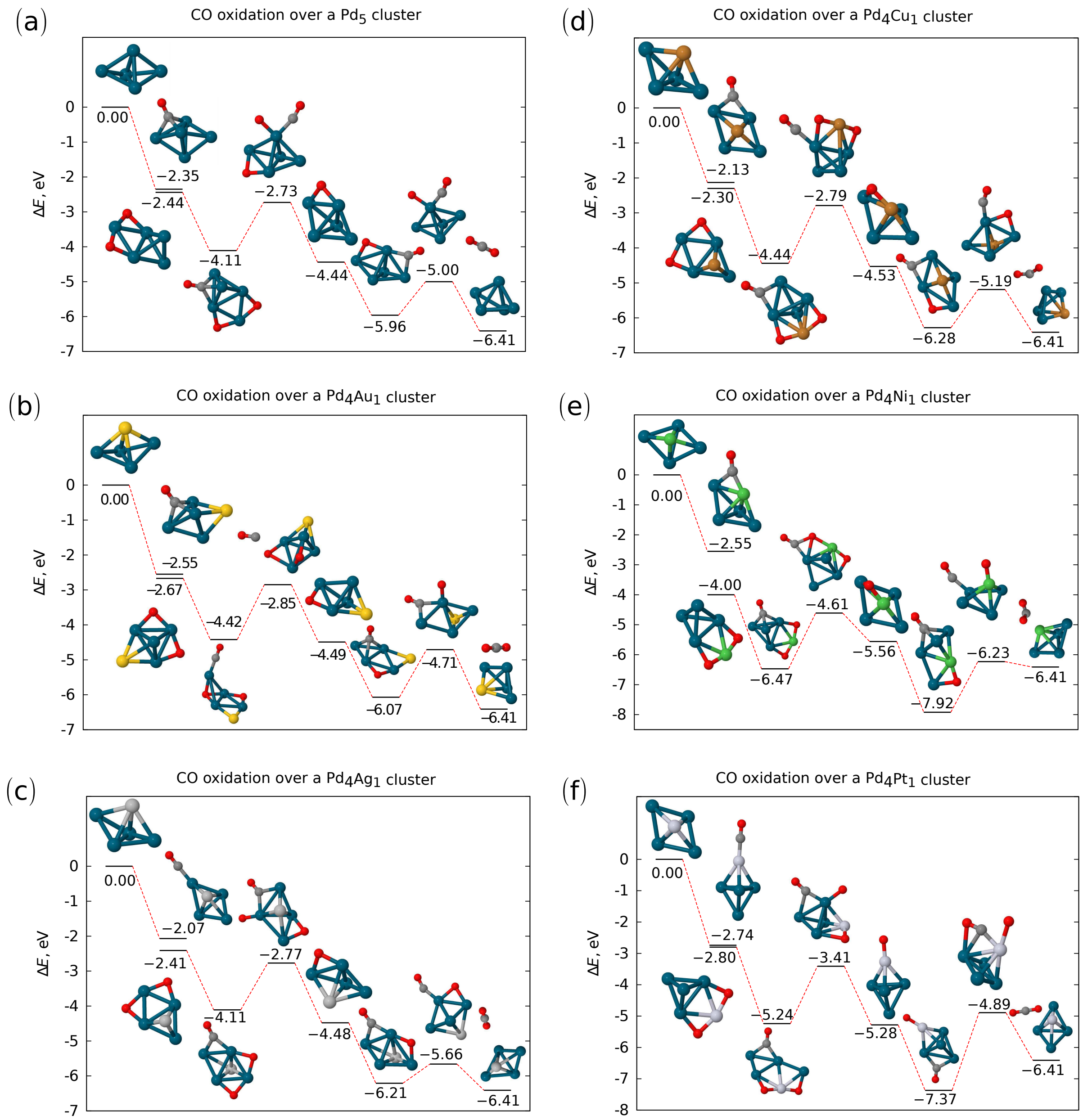

**Fig. 3** Reaction paths for $Pd_5$-based clusters: (a) $Pd_5$, (b) $Pd_4Au_1$, (c) $Pd_4Ag_1$, (d) $Pd_4Cu_1$, (e) $Pd_4Ni_1$, (f) $Pd_4Pt_1$.

cluster (Fig. 3 (a)) already exhibits the main features of a reasonably good catalyst. Firstly, every further intermediate has a relative energy which is lower than that of a previous one. Secondly, the transition state (TS) energy barriers are relatively low (1.38 eV for TS1 and 0.96 eV for TS2, see Table 1). Furthermore, both CO–$Pd_5$ and $O_2$–$Pd_5$ aggregates have very similar relative energies, which makes it more likely that both CO and $O_2$ bind to the cluster with similar probability, thus preventing the cluster from being predominantly covered by the more strongly binding species. Also noteworthy is that the global geometry optimization revealed that the geometry of the cluster itself does not change much during the reaction. In such cases, the barriers reflect just those structural changes needed for the adsorbates to react, and not any additional structural change of the cluster.

Substitution of one Pd atom with a Au atom (Fig. 3 (b))

does not change the energy profile of the reaction drastically. However, there is an important difference in how the cluster geometry changes during the reaction. In particular, the lowest-energy structures of the intermediates involving simultaneous adsorption of the CO and either an $O_2$ molecule or an O atom (i.e. $O_2CO–Pd_4Au_1$ at $-4.42$ eV and $OCO–Pd_4Au_1$ at $-6.07$ eV in Fig. 3 (b)) have a quasi-planar geometry. Such geometries turn out to be energetically most favourable, and yield lower relative energies, compared to the corresponding intermediates in the case of the pristine $Pd_5$ cluster. Therefore a substantial rearrangement of the cluster atoms is consequently needed to get to the next reaction intermediate, which probably underlies the higher energy barriers compared to $Pd_5$ (Table 1). As a result, the overall efficiency of the catalytic CO oxidation is (somewhat surprisingly) expected to be lower in the case of a gold-substituted $Pd_4Au_1$ compared to the pure $Pd_5$ cluster.

Substitution with a silver atom instead of a gold atom (Fig. 3 (c)) helps to overcome the problem of the cluster geometry changing during the reaction. As can be seen from the reaction path for the $Pd_4Ag_1$ cluster, none of the intermediates exhibit planar cluster geometries, thus the reaction involves less structural rearrangements and has lower energy barriers. Especially striking is the difference for the last step of the reaction $OCO–Pd_4M_1 \rightarrow Pd_4M_1 + CO_2$, where in the case of $Pd_4Ag_1$, only the adsorbate atoms undergo significant rearrangement; this step involves a very low energetic barrier of 0.55 eV, compared to the 1.36 eV for $Pd_4Au_1$. Thus, the $Pd_4Ag_1$ cluster appears to be a better potential catalyst for CO oxidation than $Pd_4Au_1$. Experimentally, Pd–Ag bimetallic clusters have been recently shown to be catalytically active for various reactions.[41–43]

Doping with a Cu atom also leads to reasonably good catalytic properties (Fig. 3 (d)). With copper ensuring compact structures for all the intermediates, somewhat similar to those of their Ag-substituted counterparts, the reaction path of CO oxidation over the $Pd_4Cu_1$ cluster features a second energy barrier (1.09 eV for TS2) that is lower than that of the $Pd_4Au_1$ cluster. However, the first energy barrier (1.65 eV for TS1) is comparable to that of a gold-substituted cluster. Although no substantial structural change is involved, the relatively high energy barrier can probably be attributed to the higher affinity of oxygen atoms towards copper compared to silver. It is also worth mentioning that $CO_2$ is able to physisorb on the $Pd_4Cu_1$ cluster (note that for consistency the last energy in the reaction profile in Fig. 3 (d) corresponds to the total energy release of the CO oxidation reaction, assuming $CO_2$ dissociated from the cluster). However, the small adsorption energy of 0.16 eV should not lead to any noticeable catalyst poisoning at finite temperature under actual experimental conditions. Good performance of Pd–Cu nanoalloys (although for a much larger size of $\sim$5 nm) in the process of CO oxidation has also been observed experimentally.[44]

**Table 1** Transition state energy barriers for $Pd_5$-based clsuters

| Cluster | Energy barrier, eV TS1 | Energy barrier, eV TS2 |
|---|---|---|
| $Pd_5$ | 1.38 | 0.96 |
| $Pd_4Au_1$ | 1.57 | 1.36 |
| $Pd_4Ag_1$ | 1.34 | 0.55 |
| $Pd_4Cu_1$ | 1.65 | 1.09 |
| $Pd_4Ni_1$ | 1.86 | 1.69 |
| $Pd_4Pt_1$ | 1.83 | 2.48 |

Using Pt or Ni as dopants for a Pd cluster does not lead to a potentially effective catalyst. In both cases (Fig. 3 (e) and (f)) the last $OCO–Pd_4M_1$ intermediate is more stable than the corresponding metal cluster with a newly formed $CO_2$ molecule, thus preventing the release of the second $CO_2$, and leading to poisoning of the catalyst. This tendency is also apparent from the clusters with $O_2$ and CO adsorbate molecules: both clusters have significantly higher stability for the $O_2CO–Pd_4M_1$ intermediate than either the pristine $Pd_5$ cluster, or clusters with Au, Ag, and Cu dopant atoms. This effect is most pronounced in the case of $Pd_4Ni_1$ cluster: here the affinity for $O_2$ is so high that the difference between the adsorption energies of CO and $O_2$ is as large as 1.45 eV. The $O_2CO–Pd_4Ni_1$ intermediate is also more stable than the final product of the CO oxidation reaction ($-6.47$ eV vs. $-6.41$ eV), and the $O_2CO–Pd_4Ni \rightarrow O–Pd_4Ni + CO_2$ reaction leading to the release of the first $CO_2$ is unfavourable by 0.91 eV. This can be explained by the higher mobility of the Ni $d$-electrons, where the $3d^84s^2$ configuration allows active donation towards oxidizing agents, thus preventing Ni from exhibiting properties similar to its group VIIIB noble metal neighbours.

### 4.2 Reaction paths for $Pd_6$-based clusters

Overall, $Pd_6$-based clusters exhibit similar CO oxidation behaviour. However, there are important differences caused by the increased complexity of the systems. For example, the rather strong structural distortion, which the pure $Pd_6$ cluster undergoes in TS1, changing from a $O_2CO–Pd_6$ intermediate to a $O–Pd_6$ aggregate (Fig. 4 (a)), is reflected in a rather high energy barrier of 2.10 eV (Table 2). The second transition ($OCO–Pd_6 \rightarrow Pd_6 + CO_2$) does not involve much structural change, and yields a smaller 1.35 eV barrier.

Substitution of two Pd atoms with two Au atoms enhances the catalytic activity compared to the pure $Pd_6$ cluster (Fig. 4 (b)). Unlike in the case of the singly-doped $Pd_4Au_1$ cluster, all the $Pd_4Au_2$ intermediates are non-planar. Although somewhat counterintuitive, the increased fraction of gold in the doubly-doped cluster compared to its singly-doped counterpart does not lead to an increased probability of forming planar ground-state structures for the intermediates. However, a noticeable rearrangement of the cluster's geometry is required during the first transition, which has a rather high energy barrier of 1.78 eV, compared to 1.57 eV for the corresponding transition in the singly-doped $Pd_4Au_1$ cluster. By contrast, the second transition exhibits a much smaller distortion and yields a somewhat lower energy barrier (1.29 eV for $Pd_4Au_2$ vs. 1.36 eV for $Pd_4Au_1$). As a result, $Pd_4Au_2$ overall shows more favourable properties as a potential catalyst than $Pd_6$.

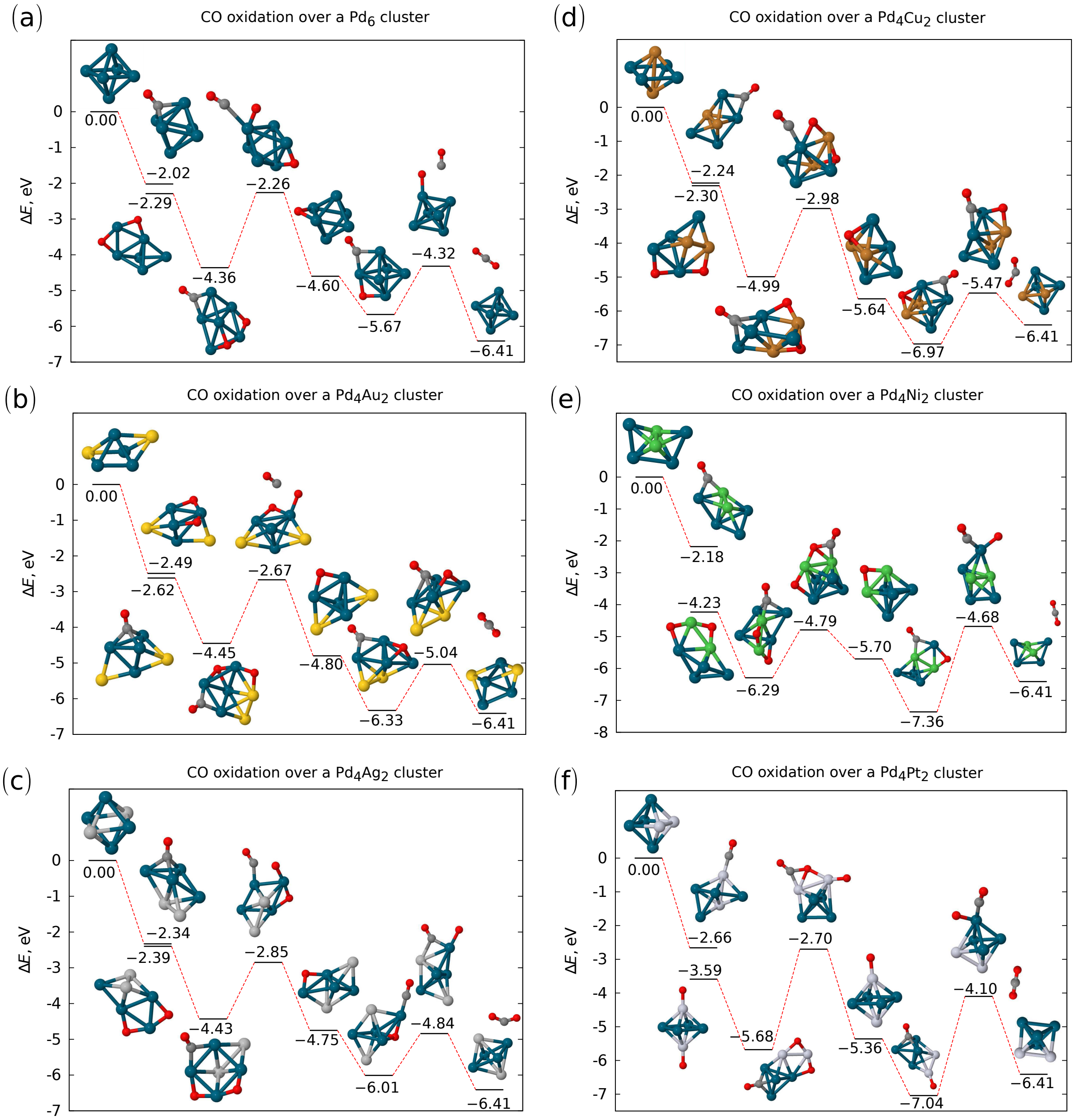

**Fig. 4** Reaction paths for $Pd_6$-based clusters: (a) $Pd_6$, (b) $Pd_4Au_2$, (c) $Pd_4Ag_2$, (d) $Pd_4Cu_2$, (e) $Pd_4Ni_2$, (f) $Pd_4Pt_2$.

Silver dopants further increase the efficiency of the doubly-doped $Pd_4Ag_2$ cluster catalyst (Fig. 4 (c)). Smaller structural distortions are reflected in the decreased values of transition state energy barriers of 1.58 eV for TS1, and 1.17 eV for TS2. Although the properties of $Pd_4Ag_2$ are thus potentially better than those of either $Pd_4Au_2$ or $Pd_6$, the $Pd_4Ag_2$ has barriers

**Table 2** Transition state energy barriers for $Pd_6$-based clsuters

| Cluster | Energy barrier, eV | |
|---|---|---|
| | TS1 | TS2 |
| $Pd_6$ | 2.10 | 1.35 |
| $Pd_4Au_2$ | 1.78 | 1.29 |
| $Pd_4Ag_2$ | 1.58 | 1.17 |
| $Pd_4Cu_2$ | 2.01 | 1.50 |
| $Pd_4Ni_2$ | 1.50 | 2.68 |
| $Pd_4Pt_2$ | 2.98 | 2.94 |

that are larger than both its smaller $Pd_4Au_1$ counterpart, and the $Pd_5$ cluster.

The Cu-substituted cluster (Fig. 4 (d)) exhibits energy barriers similar to those of the pure $Pd_6$ cluster. However, the last OCO–$Pd_4Cu_2$ intermediate is 0.56 eV more stable than a recovered catalyst cluster with a second $CO_2$ molecule produced. Moreover, physisorption of $CO_2$ that was noted earlier for the singly-copper-doped cluster is even more pronounced here: the 0.38 eV adsorption energy for $Pd_4Cu_2$ is a sign of chemical bond formation.

As is also the case for $Pd_4M_1$ clusters, Pt and Ni dopants do not assist the formation of a potentially effective $Pd_4M_2$ catalysts (Fig. 4 (e) and (f)). Addition of a second Ni atom enhances the affinity towards oxygen atoms even further: the difference between the adsorption energies of $O_2$ and CO is as large as 2.05 eV. This affinity to oxygen is also reflected in the value of the second transition state barrier: 2.68 eV obviously makes the reaction unfeasible. In the case of $Pd_4Pt_2$, additional Pt atoms cause very strong distortions to the cluster geometry, which is reflected in very high energy barriers of almost 3 eV for both transition states.

It should be pointed out that that many of the higher energy barriers are associated with transitions between intermediates with distinctively different cluster structures. This can be indicative of the difference in chemical nature of Pd clusters doped with different metals, or the increased complexity of the system, e.g. $Pd_4M_2$ compared to $Pd_4M_1$. The key question, however, is whether there are alternative pathways that lead to an overall lower reaction barrier, e.g. a multi-step process where a cluster might first change its geometry, and then the reaction occur. Such pathways may be missed by our single-path NEB calculations.

We suggest that such methodological limitations will not change our conclusion that the Ni- and Pt-doped clusters (and also $Pd_4Cu_2$) are unlikely to be good catalysts, since the unfavourable poisoning of these clusters with $CO_2$ is indepenent of whether there are alternative lower-energy paths for the reactions. For the other nanoalloys it is desirable to consider different possible reaction paths. This problem is addressed in section 4.4.

### 4.3 Electronic structure analysis of Au- and Ag-doped clusters

Comparison of the CO oxidation efficiency on $Pd_4M_1$ and $Pd_4M_2$ clusters reveals that doping can be an effective way of tuning the catalytic properties of Pd clusters. Thus, doping with silver lowers the energy barriers in both cases. The behaviour of gold dopants is, however, more complex. While $Pd_4Au_2$ exhibits potentially better catalytic properties than its pristine $Pd_6$ counterpart, the smaller $Pd_4Au_1$ yields higher transition state energy barriers due to the observed tendency to form planar intermediate structures, which leads to significant geometric rearrangement during the reaction steps.

In order to explain the observed differences between Au and Ag dopants, and also to look into the nature of chemical bonding of both CO and $O_2$ towards catalytic clusters, we calculated density of states diagrams for the $Pd_4Au_1$ and $Pd_4Ag_1$ cluster. As can be seen in Fig. 5, stronger bonding between the dopant and Pd, as well as strong binding of the adsorbates is observed in the case of the silver dopant. Kohn-Sham levels of silver and palladium show pronounced hybridization with each other in the region between $-10.0$ and $-7.5$ eV (Fig. 5 (b)). Moreover, a strong contribution from the adsorbates is also observed in this region. While a smaller interaction is also observed between palladium and gold (Fig. 5 (a)), no strong hybridization with the adsorbates is observed.

Insets in Fig. 5 depict the highest occupied molecular orbitals (HOMO) of $Pd_4Au_1$ and $Pd_4Ag_1$, which allows the behaviour of the most chemically relevant Kohn-Sham states to be directly examined. Since both clusters are triplets in their ground state, the HOMOs correspond to the highest occupied orbitals in the alpha spin channel. The HOMO of $Pd_4Au_1$ lies at $-5.39$ eV with respect to the vacuum level, while the HOMO of $Pd_4Ag_1$ is situated at $-5.16$ eV. The differences between the dopants are best observed in the case of CO dopant adsorption. As was also observed in a recent theoretical study,[21] the bonding between the CO molecule and $Pd_4Au_1$ is mostly of the $s$-$p$ $\sigma$-type, with a large fraction of electron density on the CO molecule. In the case of a silver atom dopant, however, a more complicated $d$-character is observed, indicating stronger interaction of the CO molecule with the cluster. The C$-$O bond length in the case of $Pd_4Ag_1$ is increased to 1.19 Å, compared to 1.16 Å in $Pd_4Au_1$, and 1.14 Å in equilibrium), which suggests that CO is better activated by the silver-doped cluster.

Overall, weaker adsorbate-cluster interaction, and a smaller degree of metal-metal hybridization for an Au dopant indicates that doping with Au reduces the reactivity of the cluster, compared to the addition of Ag. This renders silver as a potentially better dopant for the design of novel Pd-based bimetallic catalysts for CO oxidation.

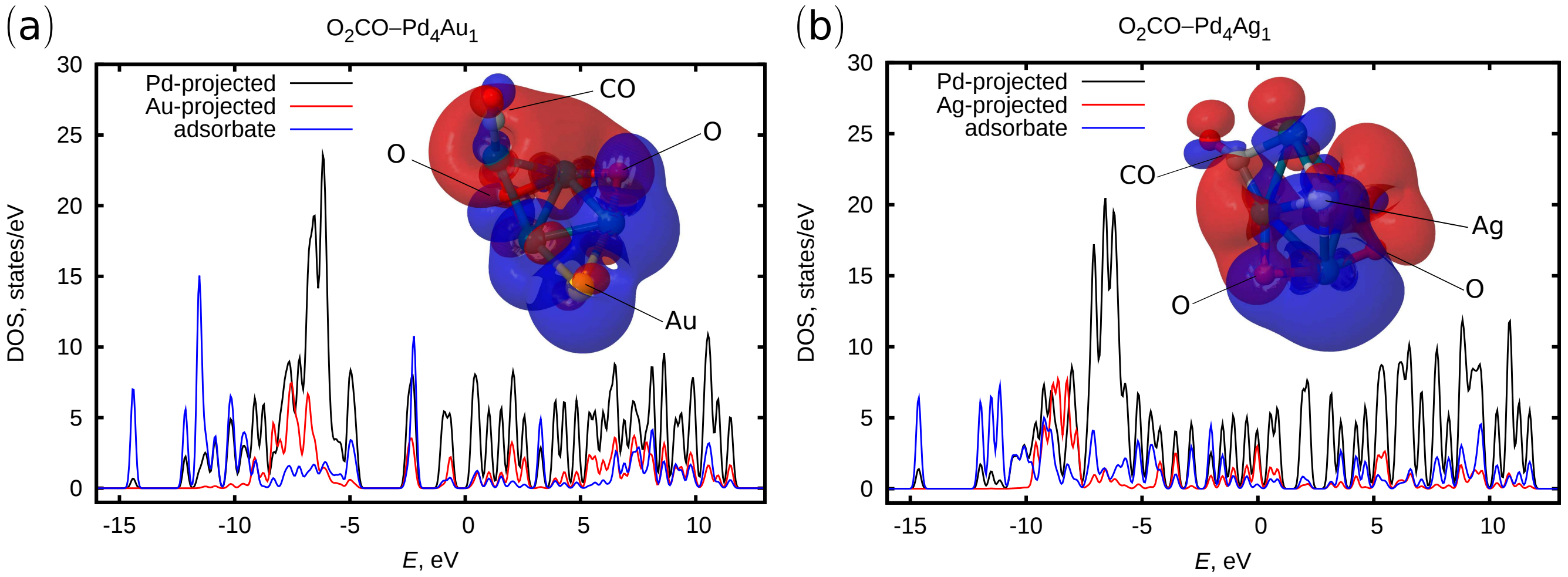

**Fig. 5** Electronic densities of states projected on Pd (black line), the dopant atom (red line), and adsorbate molecules (blue line) for (a) $O_2CO–Pd_4Au_1$, and (b) $O_2CO–Pd_4Ag_1$. The vacuum level is used as the zero reference. The insets show the highest occupied molecular orbital (HOMO) of the corresponding cluster.

### 4.4 Relative TS barriers for reaction vs. rearrangement

Calculation of the reaction paths based on the globally optimal structures for each intermediate assumes that adopting the ground-state structure of each intermediate is kinetically more likely than the immediate reaction of the adsorbate molecules on the surface of an unperturbed cluster. To explore whether this assumption is reasonable, we have chosen two systems for which pronounced changes in cluster geometry were observed along the reaction pathway, namely $Pd_4Au_1$ and $Pd_4Pt_2$. For each intermediate we then analyse low-lying isomers identified during the global optimization run, and choose the one corresponding to the least perturbation of the cluster geometry relative to the ground-state structure of the individual cluster. We then construct the reaction path diagram for such an "alternative" or "unperturbed" reaction path, and compare it to the "ground-state-based" reaction path diagram created on the basis of the globally optimal intermediates. Additionally, we also analyse the energy barriers for the reconfiguration of each reaction intermediate structure from the "unperturbed" to its ground-state geometry in order to understand whether such reconfiguration is likely to take place during the reaction.

As was suggested by the analysis of the ground-state-based reaction path, reconfigurations of the $Pd_4Au_1$ reaction intermediates are reflected in a slight increase in the transition state energy barriers for such singly-doped cluster compared to the pristine $Pd_5$ cluster. To investigate whether the "alternative" reaction path leads to lower energy barriers, in Fig. 6 we plot both the energy levels of the "alternative" and the ground-state-based reaction paths. As can be seen from the Fig. 6, all "unperturbed" intermediates are naturally less stable than the ground-state structures. Thus, at the first step of the reaction, the $CO–Pd_4Au_1$ aggregate is 0.36 eV less stable than the global minimum, while $O_2–Pd_4Au_1$ lies 0.62 eV higher than its globally optimal counterpart. However, the reconfiguration of these structures into the global minima is likely: the corresponding downhill energy barriers are only 0.06 eV in both cases (Fig. 7).

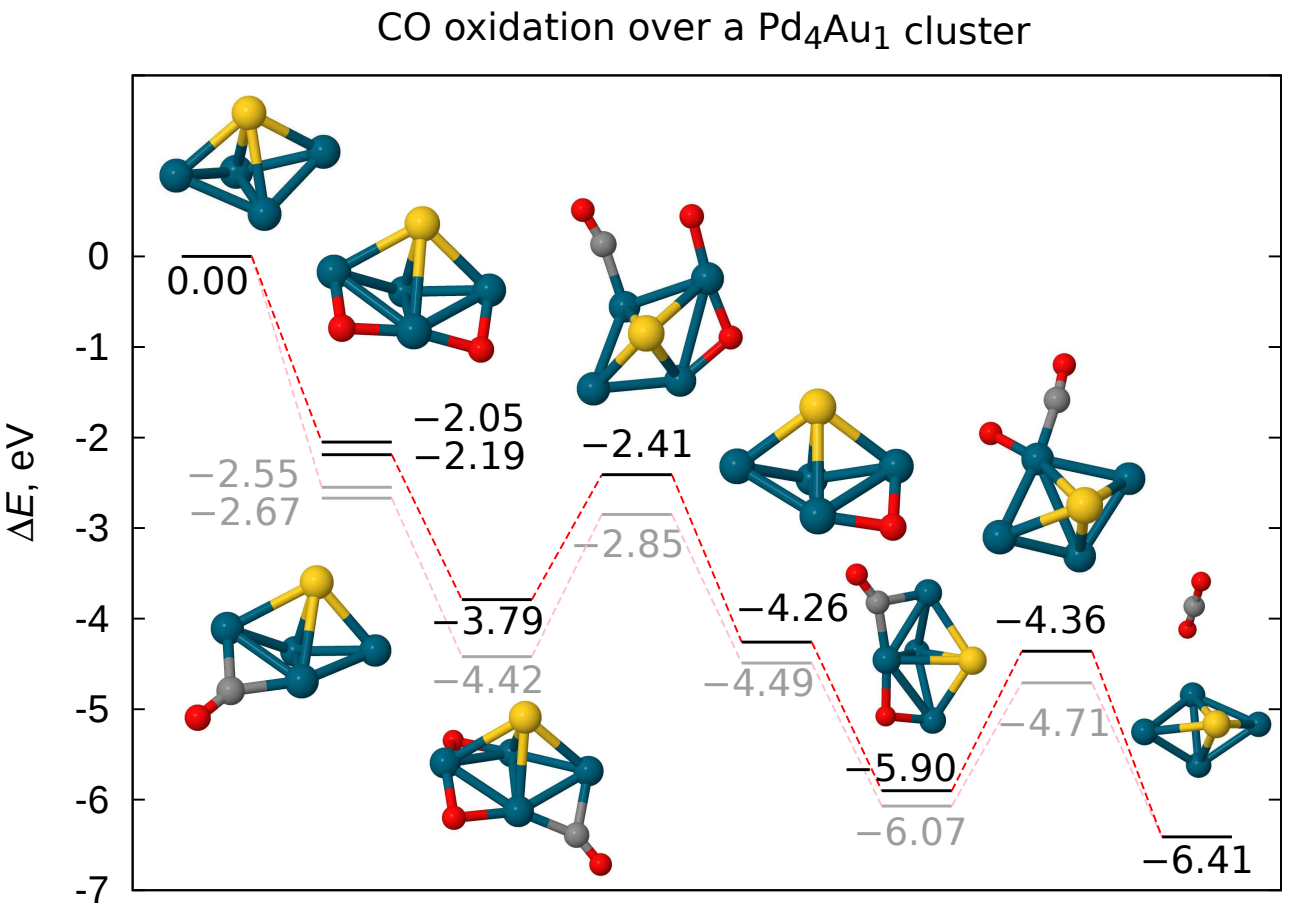

**Fig. 6** Possible reaction paths of the CO oxidation over $Pd_4Au_1$: "alternative" reaction path (black lines) compared to the ground-state-based reaction path (grey lines). Insets correspond to the "unperturbed" intermediates.

The energetic difference between the "unperturbed" and the ground-state configurations of the $O_2CO–Pd_4Au_1$ is 0.63 eV. However, reconfiguration between these quite dif-ferent structures is energetically expensive: a 2.25 eV barrier needs to be overcome. Although the first reaction barrier ($O_2CO$–$Pd_4Au_1$ $\rightarrow$ O–$Pd_4Au_1$ + $CO_2$) is slightly lower in the case of the “alternative” reaction path (1.38 eV vs. 1.57 eV), the transition state involved is 0.44 eV higher in energy than the transition state for the ground-state-based path, and so it is still likely to be more favourable to proceed along the ground-state-based path.

The next intermediate, O–$Pd_4Au_1$, lies 0.23 eV higher in energy than its ground-state structure with a downhill energy barrier of a mere 0.03 eV, which makes the formation of the global minimum configuration very likely. The two isomers of the final intermediate, OCO–$Pd_4Au_1$, lie very close to each other in energy: there is only 0.17 eV difference between the two. Although the transition to the global minimum requires a rather high barrier of 1.52 eV to be surmounted, which makes the transition less favourable. The “alternative” OCO–$Pd_4Au_1$ isomer has a higher energy barrier for the CO oxidation reaction compared to the ground-state-based path, namely 1.54 eV compared to 1.36 eV.

For the CO–$Pd_4Au_1$, $O_2$–$Pd_4Au_1$, and O–$Pd_4Au_1$ intermediates the reconfiguration required for a transition to the corresponding global minima is a simple diamond-square-diamond process between trigonal pyramidal isomers with the Au atom in apical and equatorial positions. This process facilitates an easy transition towards the more energetically favourable ground-state-based reaction path at the early stage of the reaction. By contrast, in the case of the $O_2CO$–$Pd_4Au_1$ and OCO–$Pd_4Au_1$ intermediates, the transition between an “unperturbed” cluster and the corresponding global minimum involves transformations between a non-planar geometry and a planar one, which is energetically costly, and which may make the escape from the initially chosen pathway unlikely.

In the case of $Pd_4Pt_2$, the ground-state-based reaction path features very high energy barriers, and leads to the poisoning of the catalyst due to high affinity of Pd–Pt clusters towards oxygen. Can the “alternative” reaction path based on the unperturbed pure cluster geometry be more energetically favourable? As can be seen from the Fig. 8, perturbation of the structure is a natural result of the interaction of the nanoalloy cluster with the adsorbate molecules, especially the $O_2$ molecule: both first reaction intermediates, i.e. CO–$Pd_4Pt_2$ and $O_2$–$Pd_4Pt_2$ are 0.63 eV and 1.60 eV less stable than their ground-state structures.

Such a significant energetic gain from following the ground-state-based reaction path is also conserved in the case of the second reaction intermediate $O_2CO$–$Pd_4Pt_2$ (1.63 eV less stable than the global minimum) and the first transition state TS1 ($O_2CO$–$Pd_4Pt_2$ $\rightarrow$ O–$Pd_4Pt_2$ + $CO_2$), which lies 2.19 eV higher in energy than its ground-state-based path counterpart. This yields a very high energy barrier of 3.54 eV. The resulting O–$Pd_4Pt_2$ intermediate is in turn significantly

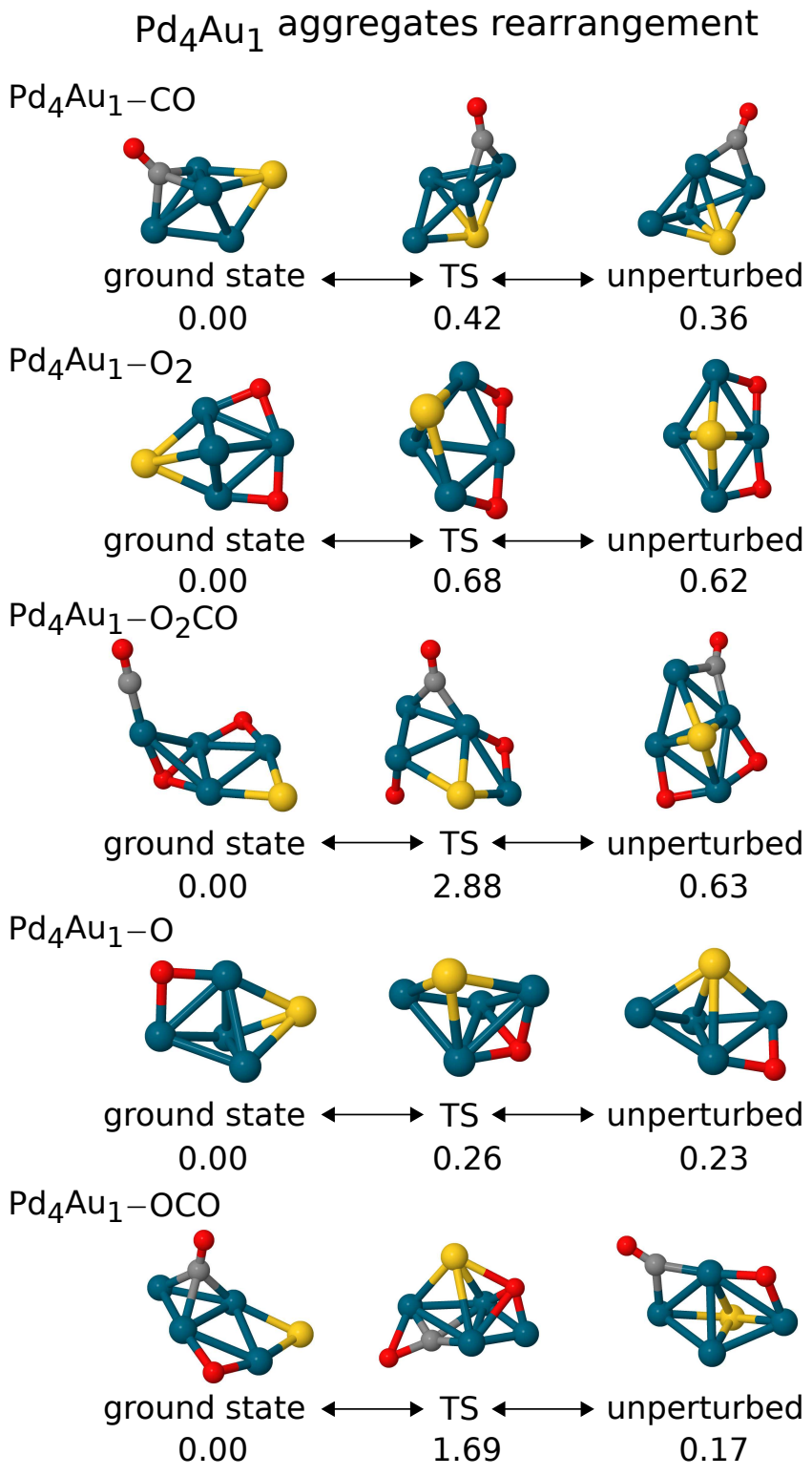

**Fig. 7** $Pd_4Au_1$ intermediate aggregates reconfiguration barriers.

less stable than its previously identified ground-state isomer (1.41 eV higher in energy).

Upon adsorption of the second CO molecule, O–$Pd_4Pt_2$ turns into OCO–$Pd_4Pt_2$, which in this case is only slightly less stable than the ground-state structure (0.35 eV difference). However, this energy is already lower than the total effect of the reaction by 0.28 eV, which again means poisoning of the catalyst, as was also observed in the case of the ground-state-based reaction path.

The second reaction transition state energy barrier turns out to be significantly lower for the “alternative” reaction path (1.07 eV vs 2.94 eV), as only adsorbate molecules undergo reconfiguration during the last reaction step (OCO–$Pd_4Pt_2$ $\rightarrow$ $Pd_4Pt_2$ + $CO_2$). However, both the prohibitively high first transition state energy barrier, and the observed poisoning of the catalyst are common to both reaction pathways and do not allow the success of the catalytic process. The energy barriers associated with the reconfiguration between the intermediates of the ground-state-based and the alternative reaction paths are presented in the supplemental material.[37]

The above results support the initially proposed significant role of the geometric configurations of the reaction intermediates on the observed transition state barriers. In the case of $Pd_4Au_1$, the ground-state-based reaction path is a reasonable assumption, as the corresponding intermediate structures are more stable, and the transition between the higher lying "unperturbed" isomers and their ground-state counterparts are kinetically likely for the first two reaction steps. In the case of $Pd_4Pt_2$, much less stable intermediate geometries still lead to a significant distortion of the structure at the stage of the release of the first $CO_2$ molecule, thus yielding an increased transition state energy barrier. Moreover, the final reaction intermediate is still lower in energy than the reaction products, which corresponds to the poisoning of the catalyst. Thus choosing ground-state configurations of the reaction intermediates appear to yield energetically reasonable reaction paths for the considered systems.

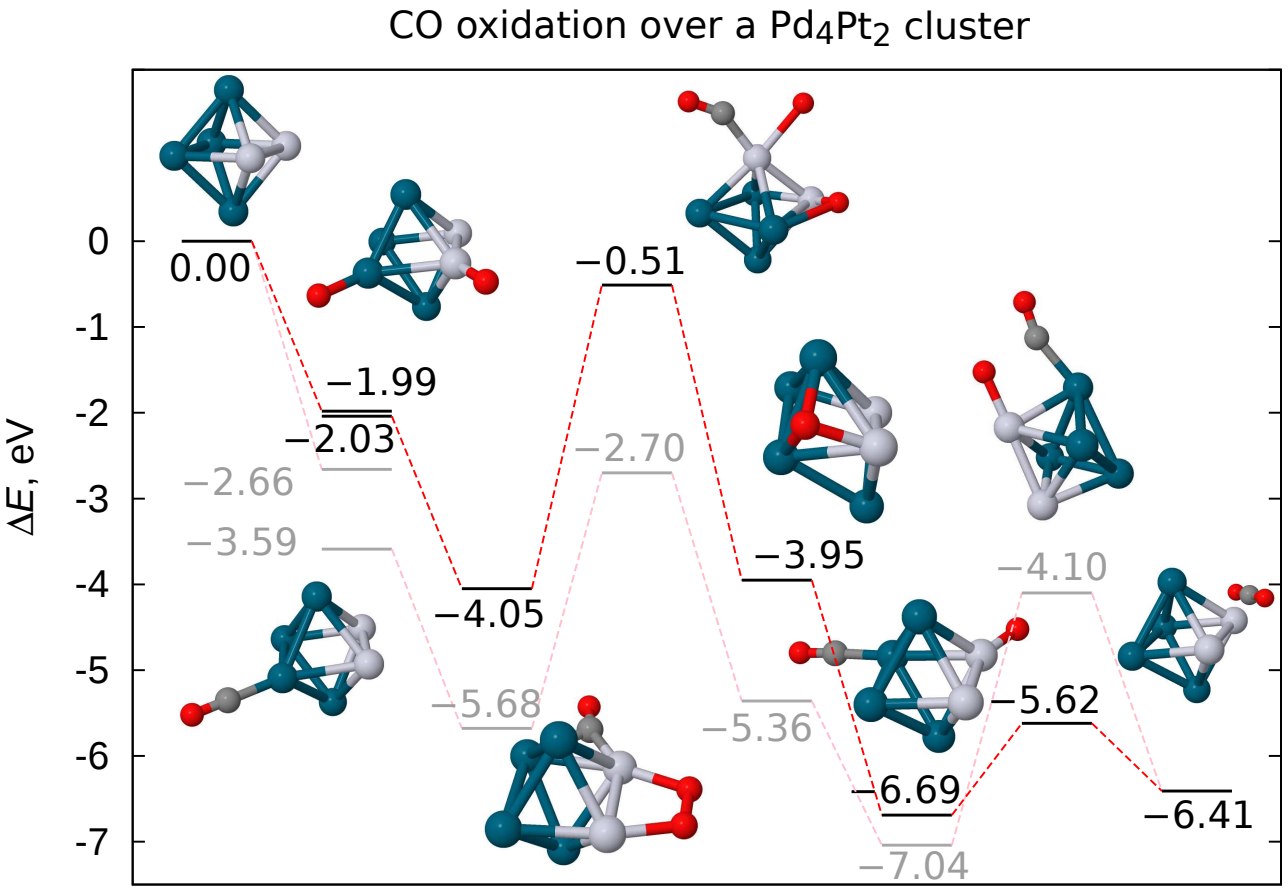

**Fig. 8** Possible reaction paths of the CO oxidation over $Pd_4Pt_2$: "alternative" reaction path (black lines) compared to the ground-state-based reaction path (grey lines). Insets correspond to the "unperturbed" intermediates.

Notwithstanding, the example of $Pd_4Pt_2$ illustrates the complexity of the configurational space that needs to be sampled to arrive at a definitive conclusion, and hence the potential pitfalls of our theoretical approach. As already mentioned above, these include the danger of picking sub-optimal configurations for the reaction intermediates, and a potential tendency of the employed transition state search method to miss more complicated multi-step pathways connecting two intermediates that might potentially involve smaller barriers.

## 5 Conclusions

In summary, we have systematically studied the influence of different transition metal dopants on the catalytic activity of Pd-based clusters in the reaction of CO oxidation. Unbiased DFT-based basin-hopping global geometry optimization reveals that the transition state energy barriers of the reaction are very sensitive to the geometric configurations of the intermediate structures. For example, substituting a Pd atom in the $Pd_5$ cluster with a Au atom leads to the increase of both transition barriers due to the tendency of gold-doped clusters to form planar structures. Noticeable rearrangement of the cluster geometry is thus needed to proceed along the reaction path, which is associated with substantial energetic costs. On the other hand, doping with silver facilitates the formation of compact structures, and helps to lower the energy barriers. Electronic structure analysis indicates that gold and silver dopants exhibit different chemical behaviour: silver atoms allow stronger hybridization with the electronic levels of both the Pd atoms in Pd cluster and the adsorbate molecules, thus creating more favourable conditions for catalytic activation of such molecules. This renders silver a potentially better dopant for the CO oxidation reaction than previously suggested gold.

Intriguingly, copper-doped nanoalloys show a potential for catalytic activity that is similar to that of the pristine palladium clusters. However, the higher affinity of oxygen for copper eventually leads to the poisoning of the catalyst in the case of $Pd_4Cu_2$. Bimetallic clusters with Ni and Pt do not exhibit favourable catalytic properties for CO oxidation due to high affinity for oxygen (Ni) and substantial structural rearrangements between the reaction steps (Pt).

Size effects might also play an important role in such bimetallic systems. Our calculations indicate that the addition of the second dopant atom increases the complexity of the system, and makes stronger geometric distortions between the reaction steps more likely, which is in turn linked to an increase in the transition state energy barriers. As a result, for both Au and Ag dopants the smaller $Pd_4M_1$ clusters appear to have better catalytic properties than the larger $Pd_4M_2$ ones. From the methodological perspective, one has to take into account that the increased configurational complexity also raises the cost of a thorough sampling of the potential energy surface, thus making it less likely that we have found the lowest-energy pathway and making it more important that additional calculations considering alternative reaction paths are undertaken.

Investigation of alternative reaction paths for the examples of $Pd_4Au_1$ and $Pd_4Pt_2$ clusters revealed that the ground-state-based reaction path, where each reaction intermediate is found as a result of the global geometry optimization, is likely to lead to a pathway with lower-energy transition states compared to the "unperturbed" reaction path, where each intermediate is represented by a low-lying isomer corresponding to the least perturbation of the cluster geometry compared to the ground state of the bare cluster. However, this is not guaranteed, particularly if there are large changes in geometry between the global optima for the reactants and products of a catalytic step. In the absence of a rigorous study of all feasible reaction paths and their possible intersections, which is prohibitively expensive for the number of systems considered here, relying on global geometry optimization for finding the configurations of the reaction intermediates appears to be a reasonable choice for constructing an energetically favourable reaction path.

Overall, the study reflects the rich diversity of bimetallic cluster structures, and high tunability of their geometric and electronic properties with a choice of a transition metal dopant, as well as the cluster size and composition. This renders such bimetallic nanoalloys as ideal candidates for the design of novel catalytic materials. We would also like to emphasize the methodological challenge of an adequate description of the configurational space of bimetallic clusters, which makes a rigorous investigation of all possible reaction paths at the electronic structure theory level unfeasible. The task ideally calls for the development of general, reliable, and computationally efficient methods of describing the potential energy surface of bimetallic clusters and adsorbate molecules, perhaps based on empirical potentials, but such a task is of course very challenging.

## Acknowledgements

Funding within the EPSRC project No. EP/J011185/1 "TOUCAN: Towards an Understanding of Catalysis on Nanoalloys" is gratefully acknowledged. All calculations were carried out using the ARCHER UK National Supercomputing Service (www.archer.ac.uk).

## References

1 U. Heiz and E. L. Bullock, *J. Mater. Chem.*, 2004, **14**, 564–577.
2 R. Ferrando, J. Jellinek and R. L. Johnston, *Chem. Rev.*, 2008, **108**, 845–910.
3 H. Ye and R. M. Crooks, *J. Am. Chem. Soc.*, 2007, **129**, 3627–3633.
4 A. Groß, *Top. Catal.*, 2006, **37**, 29–39.
5 W. Tang, L. Zhang and G. Henkelman, *J. Phys. Chem. Lett.*, 2011, **2**, 1328–1331.
6 D. Cheng and W. Wang, *Nanoscale*, 2012, **4**, 2408–2415.
7 X. Liu, C. Meng and Y. Han, *J. Phys. Chem. C*, 2013, **117**, 1350–1357.
8 P. C. Jennings, H. A. Aleksandrov, K. M. Neyman and R. L. Johnston, *Nanoscale*, 2014, **6**, 1153–1165.
9 S. M. Lang, A. Frank and T. M. Bernhardt, *Int. J. Mass Spectrom.*, 2013, **354–355**, 365–371.
10 J. Xiang, P. Li, H. Chong, L. Feng, F. Fu, Z. Wang, S. Zhang and M. Zhu, *Nano Research*, 2014, **7**, 1337–1343.
11 X. Jurvilliers, R. Schneider, Y. Fort and J. Ghanbaja, *Appl. Organomet. Chem.*, 2001, **15**, 744–748.
12 T. Weidlich and L. Prokeš, *Cent. Eur. J. Chem.*, 2011, **9**, 590–597.
13 J. Feng, C. Ma, P. J. Miedziak, J. K. Edwards, G. L. Brett, D. Li, Y. Du, D. J. Morgan and G. J. Hutchings, *Dalton Trans.*, 2013, **42**, 14498–14508.
14 A. S. K. Hashmi and G. J. Hutchings, *Angew. Chem. Int. Ed.*, 2006, **45**, 7896–7936.
15 S.-L. Peng, L.-Y. Gan, R.-Y. Tian and Y.-J. Zhao, *Comput. Theor. Chem.*, 2011, **977**, 62 – 68.
16 P. S. West, R. L. Johnston, G. Barcaro and A. Fortunelli, *Eur. Phys. J. D*, 2013, **67**, 165.
17 M. Nößler, R. Mitrić and V. Bonačić-Koutecký, *J. Phys. Chem. C*, 2012, **116**, 11570–11574.
18 L. Zhang, H. Y. Kim and G. Henkelman, *J. Phys. Chem. Lett.*, 2013, **4**, 2943–2947.
19 S. Shan, V. Petkov, L. Yang, J. Luo, P. Joseph, D. Mayzel, B. Prasai, L. Wang, M. Engelhard and C.-J. Zhong, *J. Am. Chem. Soc.*, 2014, **136**, 7140–7151.
20 *Computational Catalysis*, ed. A. Asthagiri and M. J. Janik, The Royal Society of Chemistry, 2014, pp. 1–266.
21 J. Zhang and A. N. Alexandrova, *J. Phys. Chem. Lett.*, 2013, **4**, 2250–2255.
22 N. Guo, R. Lu, S. Liu, G. W. Ho and C. Zhang, *J. Phys. Chem. C*, 2014, **118**, 21038–21041.
23 S. J. Clark, M. D. Segall, C. J. Pickard, P. J. Hasnip, M. J. Probert, K. Refson and M. Payne, *Z. Kristall.*, 2005, **220**, 567–570.
24 J. P. Perdew, K. Burke and M. Ernzerhof, *Phys. Rev. Lett.*, 1996, **77**, 3865–3868.
25 J. Nocedal and S. J. Wright, *Numerical Optimization*, Springer US, 2006.
26 D. J. Wales and J. P. K. Doye, *J. Phys. Chem. A*, 1997, **101**, 5111–5116.
27 D. J. Wales, J. P. K. Doye, M. A. Miller, P. N. Mortenson and T. R. Walsh, *Adv. Chem. Phys.*, 2000, **115**, 1–111.
28 S. R. Bahn and K. W. Jacobsen, *Comput. Sci. Eng.*, 2002, **4**, 56–66.
29 G. Henkelman, B. P. Uberuaga and H. Jónsson, *J. Chem. Phys.*, 2000, **113**, 9901–9904.
30 *CRC Handbook of Thermophysical and Thermochemical Data*, ed. D. R. Lide and H. V. Kehiaian, CRC Press, Boca Raton, 1994.
31 H. Häkkinen and U. Landman, *Phys. Rev. B*, 2000, **62**, R2287–R2290.
32 H. Häkkinen, B. Yoon, U. Landman, X. Li, H.-J. Zhai and L.-S. Wang, *J. Phys. Chem. A*, 2003, **107**, 6168–6175.
33 A. Castro, M. A. L. Marques, A. H. Romero, M. J. T. Oliveira and A. Rubio, *J. Chem. Phys.*, 2008, **129**, 144110.
34 S. Lee, C. Fan, T. Wu and S. L. Anderson, *J. Am. Chem. Soc.*, 2004, **126**, 5682–5683.
35 W. E. Kaden, T. Wu, W. A. Kunkel and S. L. Anderson, *Science*, 2009, **326**, 826–829.
36 L. Li, Y. Gao, H. Li, Y. Zhao, Y. Pei, Z. Chen and X. C. Zeng, *J. Am. Chem. Soc.*, 2013, **135**, 19336–19346.
37 See supplementary material for further details on the structures and relative energies of the $Pd_{5-x}Au_x$ and $Pd_{6-x}Au_x$ clusters adsorbed on the $TiO_2$ (110) surface, and the energy barriers associated with the reconfiguration between the intermediates of the ground-state-based and the alternative reaction paths for the $Pd_4Pt_2$ cluster.
38 D. Çakir and O. Gülseren, *J. Phys. Chem. C*, 2012, **116**, 5735–5746.
39 A. S. Mazheika, T. Bredow, O. A. Ivashkevich and V. E. Matulis, *J. Phys. Chem. C*, 2012, **116**, 25274–25285.
40 W. E. Kaden, W. A. Kunkel, F. S. Roberts, M. Kane and S. L. Anderson, *J. Chem. Phys.*, 2012, **136**, 204705.
41 G. Barcaro and A. Fortunelli, *Faraday Discuss.*, 2008, **138**, 37–47.
42 Z. Yin, Y. Zhang, K. Chen, J. Li, W. Li, P. Tang, H. Zhao, Q. Zhu, X. Bao and D. Ma, *Sci. Rep.*, 2014, **4**, 4288.
43 Z. Cui, M. Yang and F. J. DiSalvo, *ACS Nano*, 2014, **8**, 6106–6113.
44 V. Abdelsayed, A. Aljarash, M. S. El-Shall, Z. A. Al Othman and A. H. Alghamdi, *Chem. Mater.*, 2009, **21**, 2825–2834.

# Supplemental Material to
# CO Oxidation Catalysed by Pd-based Bimetallic Nanoalloys

Dennis Palagin* and Jonathan P. K. Doye

## I. Structures and relative energies of the $Pd_{5-x}Au_x$ and $Pd_{6-x}Au_x$ clusters adsorbed on the $TiO_2$ (110) surface

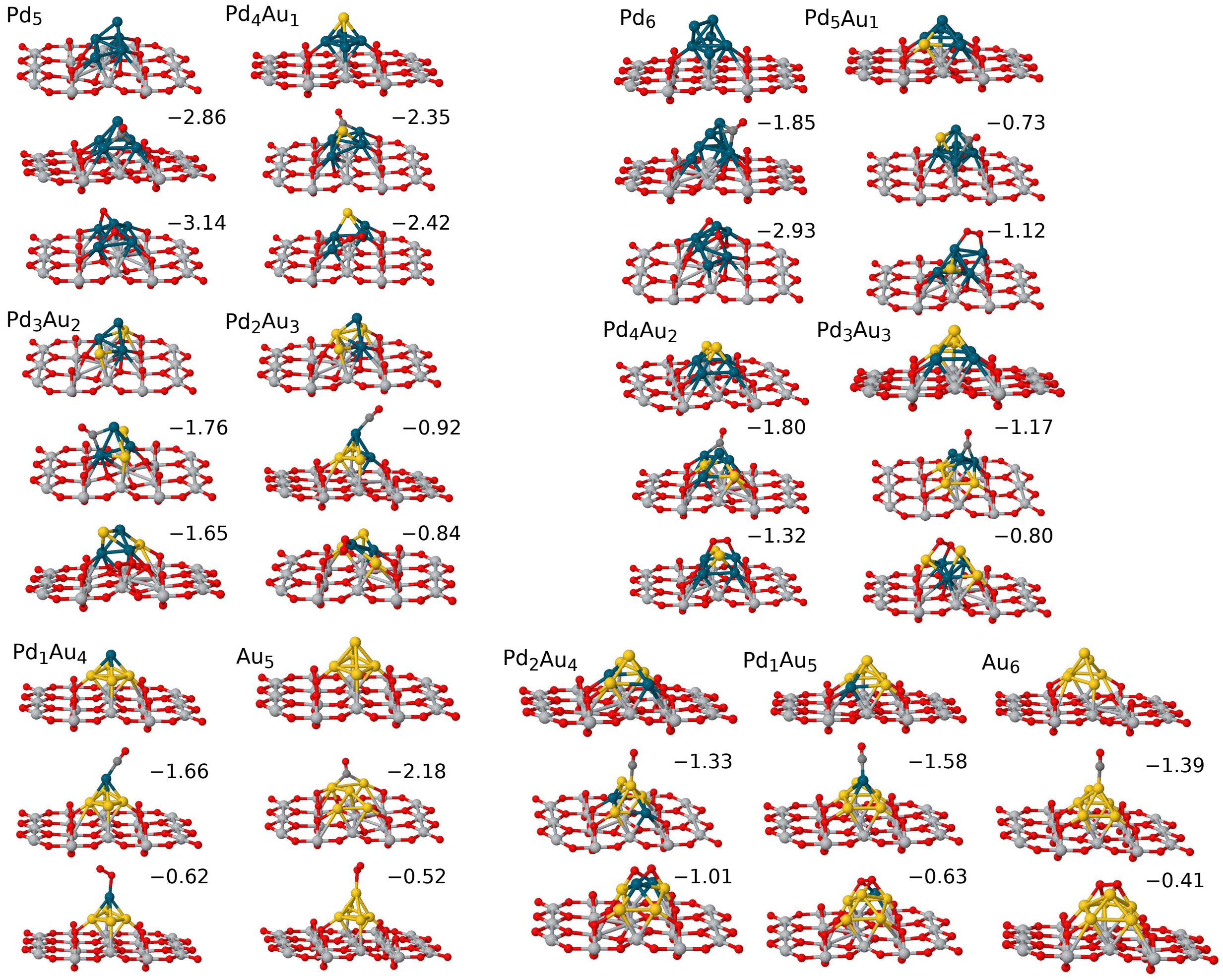

FIG. S1. Lowest-energy $Pd_{5-x}Au_x$ and $Pd_{6-x}Au_x$ clusters adsorbed on the $TiO_2$ (110) surface, and their aggregates with CO and $O_2$. Numbers indicate adsorption energies in eV.

* dennis.palagin@chem.ox.ac.uk

## II. $Pd_4Pt_2$ intermediate aggregates reconfiguration

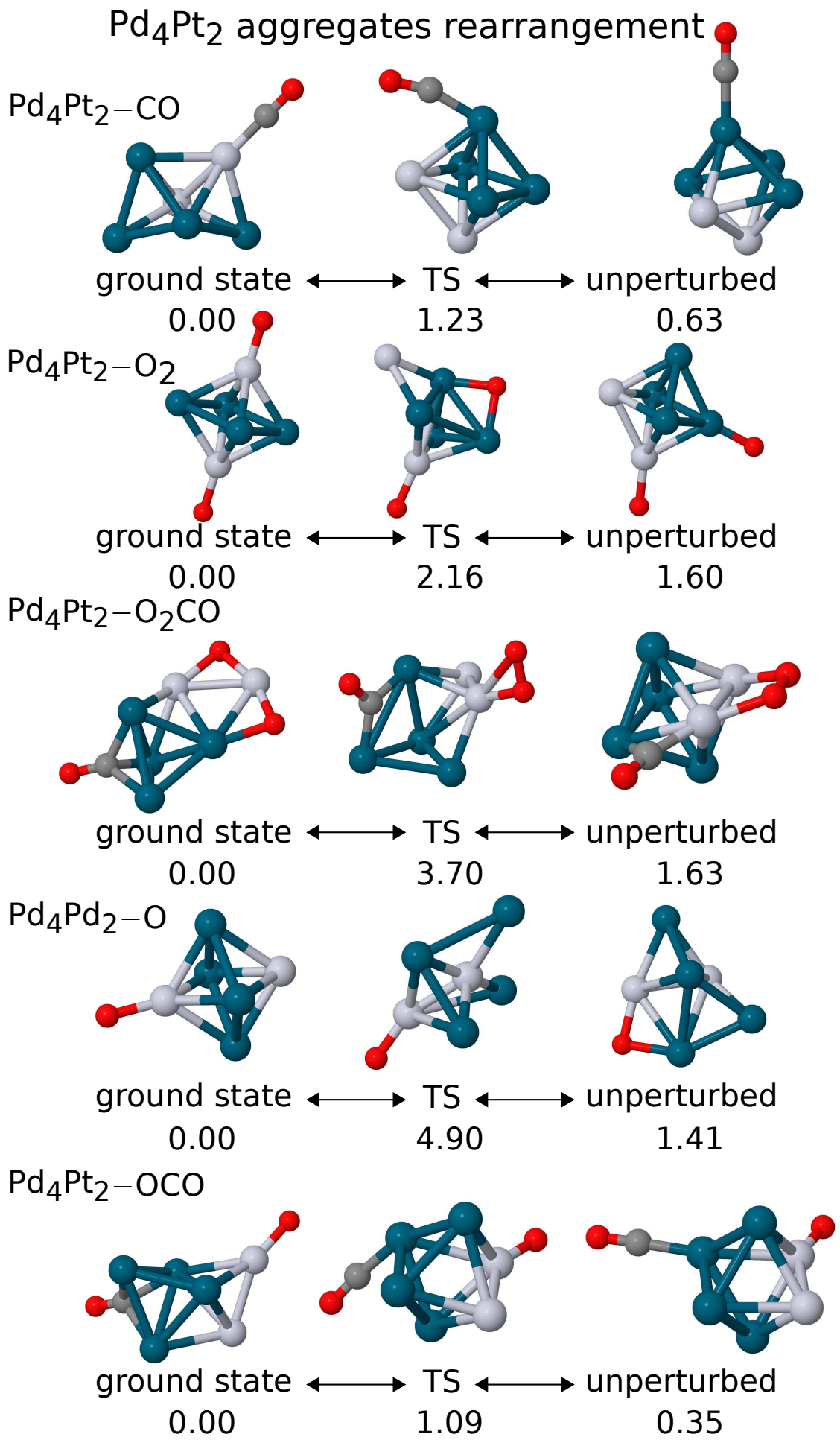

FIG. S2. $Pd_4Pt_2$ intermediate aggregates reconfiguration barriers.